\documentclass{tMOP2e}

\usepackage{epstopdf}
\usepackage{subfigure}
\usepackage{lineno}
\theoremstyle{plain}

\usepackage{color}

\theoremstyle{definition}

\theoremstyle{remark}

\begin{document}


\articletype{}

\title{\textit{Squeezing of the mechanical motion and beating 3 dB limit using dispersive optomechanical interactions } }

\author{
\name{U. Satya Sainadh\textsuperscript{a} and M. Anil Kumar\textsuperscript{b}$^{\ast}$}
\thanks{$^\ast$Corresponding authors. Email:   anilk@rri.res.in }
\affil{\textsuperscript{a}Center for Quantum Dynamics, Griffith Universiy, Brisbane, Queensland 4111, Australia;\\\textsuperscript{b} Raman Research Institute, Bangalore, India-560080.
}
}

\maketitle

\begin{abstract}
We study an optomechanical system consisting of an optical cavity and movable mirror coupled through dispersive linear optomechanical coupling (LOC) and quadratic optomechanical coupling(QOC). We work in the resolved side band limit with a high quality factor mechanical oscillator in a strong coupling regime. We show that the presence of QOC in the conventional optomechanical system (with LOC alone) modifies the mechanical oscillator's frequency and reduces the back-action effects on mechanical oscillator. As a result of this the fluctuations in mechanical oscillator can be suppressed below standard quantum limit thereby squeeze the mechanical motion of resonator. We also show that either of the quadratures can be squeezed depending on the sign of the QOC. With detailed numerical calculations and analytical approximation we show that in such systems, the 3 dB limit can be beaten. 
\end{abstract}


\section{Introduction}

An optomechanical system (Fig.~\ref{fig1}), consisting of an interaction and mutual control between movable mechanical mirror (oscillator) and optical cavity modes, is a big demand for studying quantum features at a macroscopic scale. 
 These systems were investigated \cite{braginsky,braginskybook2,braginskybook} and theoretically described \cite{cklawform} very early in the context of ponderomotive effects of light on moving mirrors. Exploiting optics to affect and study the mechanical degrees of freedom enabled physicists to conceive gravitational wave detectors on these principles \cite{caves}, where the mechanical oscillator shall be in a squeezed state. To achieve such squeezed states in mechanical oscillator, conventional optomechanical systems (with linear optomechanical coupling (LOC)) were proposed along with feeding an additional periodically modulated light fields \cite{eisert,farace} and with large cavity detuning \cite{cklaw,maquardt}. Also in such systems squeezing was predicted by parametrically driving the mechanical oscillator coupled to a microwave cavity \cite{milburn}.  Apart from the above mentioned schemes, various techniques were proposed such as quantum feedback process  \cite{ruskov,clerk,doherty,doherty-weak,vanner,vinante,pontin}, by injecting a squeezed light in to the cavity \cite{zoller,agarwal,li} and quantum reservoir engineering \cite{kronwald}. 


Recent experimental advancements based on \cite{kronwald}  in microwave domain has enabled squeezing of the mechanical oscillator's quadrature. Wollman \textit{et.al.} achieved it to a minimum variance of 0.8 times that of the ground state \cite{mechsqueezeing}, Pirkkalainen \textit{et.al.} achieved motional squeezing of a macroscopic oscillator by limiting its quantum noise to $1.1\pm 0.4$ dB below the standard quantum limit (SQL) \cite{brandt} and finally quantum fluctuations of the mechanical oscillator was reduced 20\% below the quantum noise \cite{teufel}. However an optomechanical system with dispersive LOC has fundamental limitations due to the back-action, which ultimately limits the mechanical oscillator's fluctuations to the SQL. In order to suppress fluctuations further, it is necessary to take a step beyond the standard LOC and exploit either higher-order dispersive optomechanical couplings or mechanical non-linearities. Though there have been proposals in this direction using quadratic optomechanical couplings (QOC) alone \cite{girvin,asjad} and duffing non-linearity \cite{duffing}. 
 
 Recent experiments have measured QOC in various optomechanical setups such as membrane in the middle set-up \cite{thompson,sankey,karuza,jharris}, atomic gases trapped in Fabry-Pérot cavities \cite{purdy}, microdisk-cantilever systems \cite{doolin}, microsphere-nanostring systems \cite{nanostring}, paddle nanocavities \cite{kaviani} and tunable photonic crystal Optomechanical Cavity \cite{pariso,painterarxiv}. Among all these schemes the tunable photonic crystal system enabled to yield the value of QOC of 245 Hz \cite{pariso}  and the recent advancement has pushed this limit to the order of kHz \cite{painterarxiv}. In \cite{thompson,sankey} a membrane placed at node or antinode of the cavity mode showed the presence of QOC. Further Sankey \textit{et. al.} also showed that the membrane possessed an angular degree of freedom (tilt) that allowed a smooth transition of QOC from positive to negative value. Such a control on QOC was also observed in an atom-chip based system \cite{purdy}, where the sign of QOC was correlated to the compression and expansion of atomic cloud when placed at node and anti-node of the cavity mode. Also the proposal made using hybrid optomechanical system \cite{xureb} can generate negative or positive QOC depending on the position of dielectric nano-sphere at the node or antinode of the optical cavity mode. In accordance with these advancements made in controlling the strength and nature of QOC, we present a way of beating SQL  in optomechanical systems by using both dispersive LOC and residual QOC (\textit{i.e.} $\text{LOC}\gg \text{QOC}$) together. 
 
 There has been extensive work on the generation, control and modification of the squeezed states in a harmonic oscillator due to instantaneous changes in its frequency \cite{graham,janszky,mass-frequency} or spring constant \cite{rugar}. We utilise such a similar mechanism in our system in the resolved sideband regime where the presence of QOC minimizes the back-action effects on mechanical oscillator. Using extensive numerical and analytical calculations, we not only show our approach as a useful tool to squeeze either of the quadratures but also reckon that such a system is helpful in breaking the 3 dB limit that is paramount for ultra sensitive precision measurements and some quantum information applications \cite{braunstein}. 

The paper is organised as follows: Sec. 2 presents the hamiltonian describing the system with equations of motion. It is followed by Sec. 3 consisting of linearisation of the equations of motion giving the steady state mean values and stability criteria. Section. 4 and Sec.5 presents the theoretical description of mechanical oscillator's fluctuations and standard quantum limit respectively. Later, in Sec. 6  we present the numerical results and analytical approximation to understand the effect of QOC on mechanical squeezing. Finally conclusions are given in Sec. 7.

\section{Model system and Hamiltonian}
\begin{figure}[t] \centering
 \includegraphics[scale=0.72]{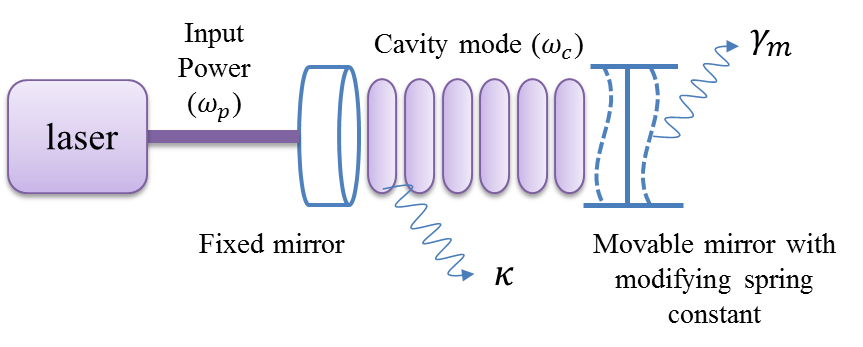}
 \caption{\label{fig1}   Schematic of a cavity optomechanical system with an optical mode supported in the cavity of line width $\kappa$ and resonance frequency $\omega_c$, driven by an external laser of frequency $\omega_p$ and power $\mathcal{P}$. The radiation pressure interacting through linear and quadratic couplings exert force on the movable mirror that softens or hardens the spring, modifying its natural frequency $\omega_m$ and damping rate $\gamma_m$.}
 \end{figure}

 We consider the optomechanical
 system of Fig. \ref{fig1}, with a single cavity mode frequency $\omega_c$ 
and cavity decay rate $\kappa$  coupled to a single mechanical mode of
 frequency $\omega_m$ driven by a strong pump field of frequency $\omega_p$. The interactions are both linear $(g_{_{l}})$ and quadratic $(g_{_{q}})$ in mechanical oscillator's displacement. The Hamiltonian 
 of the system in the laser frame is given by
 \small
 \begin{equation}
 H=\hbar\Delta \mathbf{a^\dagger a}+\frac{\hbar \omega_m }{2}(\mathbf{x}^2+\mathbf{p}^2)+\hbar g_{_{l}} \mathbf{a^\dagger a x}+\hbar g_{_{q}} \mathbf{a^\dagger a}\mathbf{x}^2+i\hbar\mathbf{\varepsilon(a^\dagger-a)} \label{a1}
 \end{equation}\normalsize
with $\Delta=\omega_c-\omega_p$ being the cavity detuning and 
$\mathbf{a (a^\dagger)}$ is the annihilation (creation) operator of the cavity mode such that $[\mathbf{a,a^\dagger} ]  = 1$. Here $\mathbf{x}$ and $\mathbf{p}$ refer
to the dimensionless position and momentum operators for the mechanical oscillator with the commutation relation as $[\mathbf{x},\mathbf{p}]=i$. While the first two terms express the free energy of the optical field and mechanical oscillator, the next two terms describe the linear and quadratic optomechanical interactions. These interactions couple the mechanical oscillator to the optical field linearly and quadratically in its  displacement.
The LOC and QOC constants are defined as  $g_{_{l}}=\frac{\partial \omega_c}{\partial x} \sqrt{\frac{\hbar}{m \omega_m}}$ and  $g_{_{q}}=\frac{\partial^2 \omega_c}{\partial x^2} \frac{\hbar}{2m \omega_m}$, respectively with $m$ being the effective mass of the oscillator. 
 The last term describes the interaction of the cavity mode with pump field amplitude 
$(\varepsilon=\sqrt{\frac{2\kappa \mathcal{P}}{\hbar \omega_p}})$, $\mathcal{P}$ being the input power of the pump field. 
In order to fully describe the dynamics of the system
  we use the Hamiltonian Eqn.(\ref{a1}) and consider the dissipation forces. The time evolution of the system operators can be derived by using the Heisenberg
equations of motion and adding the corresponding damping and noise terms yields us the following quantum Langevin
 equations:
 \small
 \begin{subequations}
 \begin{eqnarray} 
 &&\dfrac{d\mathbf{x}}{dt}=\omega_m \mathbf{p},\\
 &&\dfrac{d\mathbf{p}}{dt}=-\omega_m \mathbf{x}-g_{_{l}}\mathbf{a^\dagger a} -2g_{_{q}} \mathbf{a^\dagger a x} - \gamma_m \mathbf{p}+\xi(t),\\
 &&\dfrac{d\mathbf{a}}{dt}=-(\kappa+i\Delta) \mathbf{a} -i g_{_{l}}\mathbf{a x}-ig_{_{q}}  \mathbf{a} \mathbf{x}^2 +\varepsilon +\sqrt{2 \kappa}\mathbf{a_{in}}. \end{eqnarray}\end{subequations}\label{s1}\normalsize
Here the noise in the field and mechanical mode is described by noise operator $\mathbf{a_{in}}$ with zero mean and $\delta a_{in}(t)$ fluctuations around it and  the Brownian stochastic force described by $\xi(t)$ with zero-mean, having a damping rate $\gamma_m$, respectively. The correlation function at temperature T for these noises are  \small
\begin{subequations} \begin{eqnarray}
 &&\langle \xi(t)\xi(t')\rangle= \frac{\gamma_m}{2\pi\omega_m}\int\omega e^{-i\omega(t-t')}\left[1+\coth\left(\frac{\hbar\omega}{2k_BT} \right)\right]d\omega, \label{a13}\\
 && \langle \delta a_{in}(t)\delta a_{in}^\dagger(t')\rangle =(n_a+1)\delta (t-t'),\\
 &&\langle \delta a_{in}^\dagger(t)\delta a_{in}(t')\rangle =n_a\delta (t-t').
\end{eqnarray}\end{subequations} \normalsize where  $k_B$ is the Boltzmann constant and $n_a=\left[ \exp(\frac{\hbar\omega_c}{k_BT})-1\right]^{-1}$ is the equilibrium mean thermal photon number. At optical frequencies $\hbar \omega_c \gg k_BT$ and therefore $n_a\approx 0$.
\section{Linearisation and steady-state analysis}
We rewrite the Heisenberg operators as complex numbers  with $\langle \mathbf{O} \rangle \equiv O$, representing their respective steady state values
 with the inclusion of fluctuations around their steady state values,
 $i.e$ $O(t)=O_s+\lambda \delta O(t)$. Expanding the set of equations Eqn.(\ref{s1}) in the manner described above leads us to a set of non-linear algebraic equations for steady state values, given by\small
 \begin{subequations}\begin{eqnarray}
 &&  x_s= \frac{-g_{_{l}}|a_s|^2}{\omega_m+2 g_{_{q}}|a_s|^2} ,
\label{a4}\\
  &&p_s=0,\label{a3}\\
  &&  a_s  =\frac{\varepsilon}{ \kappa+i\left( \Delta+g_{_{l}}x_s +g_{_{q}}   (x^2)_s \right)}.
 \end{eqnarray}\label{a5}\end{subequations}\normalsize
   
  Since QOC is central to our paper we also calculate the evolution of the expectation values of $\mathbf{x}^2$, $\mathbf{p}^2$ and $\mathbf{xp+px}$ using the same method and the factorisation assumption $\langle \mathbf{O_1O_2}\rangle=\langle \mathbf{O_1}\rangle \langle\mathbf{ O_2}\rangle$ giving us,
  \small
   \begin{subequations}
 \begin{eqnarray}
 &&\dfrac{d x^2 }{dt}=\omega_m  (xp+px) ,\\
 &&\dfrac{d p^2}{dt}=-\left(\omega_m+2 g_{_{q}}a^\dagger a\right) (xp+px) -2 g_{_{l}} a^\dagger a  p-2\gamma_m  p^2  \nonumber \\&&\hspace*{1cm}+2\gamma_m(1+2n_{th}),\\
 &&\dfrac{d (xp+px) }{dt}=-2 \left(\omega_m+2 g_{_{q}} a^\dagger a \right) x^2  -2 g_{_{l}} a^\dagger a x +2 \omega_m  p^2  \nonumber\\&&\hspace*{2cm}-\gamma_m (xp+px), \nonumber\\
 \end{eqnarray} \label{a2}\end{subequations}
 \normalsize where $n_{th}=[exp \left(\frac{\hbar \omega_m}{k_B T}\right)-1]^{-1}$ is the mean thermal phonon number.
 Calculating the steady state values of these bilinear quantities in a similar way as before, we get
 \small 
 \begin{subequations}
 \begin{eqnarray}
 && (x^2)_s=\frac{g_{_{l}}^2|a_s|^4}{(\omega_m +2 g_{_{q}}|a_s|^2)^2}+\frac{\omega_m (1+2 n_{th})}{\omega_m +2 g_{_{q}}|a_s|^2},\\
 &&(p^2)_s=1+2n_{th},\\
 &&(xp+px)_s=0.\end{eqnarray}\end{subequations}\normalsize
 
Since the fluctuations are assumed to be small when compared to the steady state values, we can neglect the non-linear terms in $\lambda$. This enables us to write the linearised Langevin equations for the fluctuations from  Eqn.(\ref{s1}), in a compact form as shown below, 
 \begin{equation}\dot{u}(t)=M u(t)+\nu(t),\label{a12}
\end{equation} with column vector of fluctuations in the system being $u(t)^T= \left(\begin{matrix} \delta x(t),\delta p(t),\delta X(t),\delta P(t) \end{matrix}\right)$, column vector of noise being 
$\nu(t)^T=\left(\begin{matrix}0,\xi(t),\sqrt{2\kappa}\delta X_{in}(t),\sqrt{2\kappa}\delta P_{in}(t)\end{matrix}\right)$, using the definitions
 $\delta X \equiv \frac{\delta a+\delta a^\dagger}{\sqrt{2}}$, $\delta P \equiv \frac{\delta a-\delta a^\dagger}{\sqrt{2}i}$ and their corresponding noises $\delta X_{in}$ and $\delta P_{in}$. The matrix $M$ is given by
\small
\begin{eqnarray}M=\left(\begin{matrix} 0&\omega_m &0&0\\ -\tilde{\omega}_m& -\gamma_m& 
-\tilde{G} X_s&-\tilde{G} P_s\\ \tilde{G} P_s& 0&-\kappa  & \tilde{\Delta}\\ -\tilde{G} X_s & 0& -\tilde{\Delta}&-\kappa \end{matrix}\right) ,
\end{eqnarray}\normalsize
with  $I\equiv |a_s|^2$, $\tilde{\omega}_m \equiv \omega_m+2g_{_{q}} I$, $\tilde{\Delta}\equiv \Delta+g_{_{l}}x_s +g_{_{q}} x_s^2 $, $X_s=\frac{a_s+a_s^*}{\sqrt{2}}$, $P_s=\frac{a_s-a_s^*}{\sqrt{2}i}$ and
$\tilde{G} \equiv g_{_{l}}  +2 g_{_{q}}x_s$. 
The solutions of Eqn.(\ref{a12}) are stable only if all the eigenvalues of the matrix $M$ have negative real parts. This can be deduced by applying Routh-Hurwitz criterion \cite{routh} giving the following conditions in terms of system parameters:
\small
\begin{subequations}
\begin{eqnarray}
&&s_1 \equiv (\kappa^2+\tilde{\Delta}^2)+2 \kappa \gamma_m+\tilde{\omega}_m \omega_m >0,\\
&& s_2 \equiv (\kappa^2+\tilde{\Delta}^2)\gamma_m+2 \kappa \tilde{\omega}_m \omega_m>0,\\
&&s_3 \equiv (\kappa^2+\tilde{\Delta}^2)\tilde{\omega}_m \omega_m-\tilde{\Delta} \omega_m \tilde{G}^2 (X_s^2+P_s^2)>0,\\
&& (2 \kappa +\gamma_m) s_1 > s_2, \\
&& s_1 s_2 (2 \kappa +\gamma_m)> s_2^2+(2 \kappa +\gamma_m)^2 s_3.
\end{eqnarray}\label{a15}\end{subequations}\normalsize

We analyse the effects of QOC on this system by fixing LOC and varying QOC. In all our numerical results we scale QOC values with LOC value shown as $g_{_{q}}/g_{_{l}}$. The parameters chosen in our calculations are similar to those used in \cite{vitali}: driven by a laser of wavelength 810 nm . The mechanical oscillator  has a frequency $ \omega_m/2\pi=10$ MHz, damping rate $\gamma_m/2\pi=100$ Hz, mass $m=5$ ng and LOC, $g_{_{l}}/2\pi = 215$ Hz. In order to work in the resolved side-band limit we choose cavity damping rate as $\kappa/2\pi= 1$ MHz, and thermal bath temperature 1 mK. 
\begin{figure}[t]
\hspace*{0cm} \includegraphics[scale=0.373]{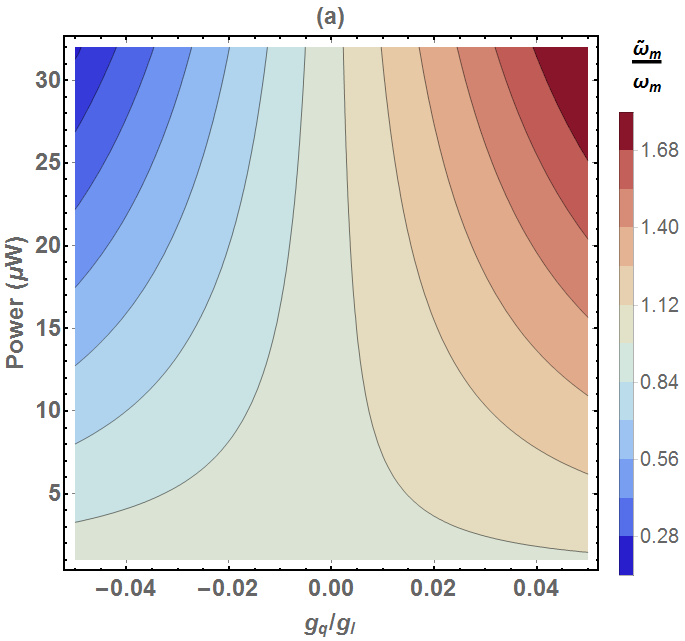}  \hspace*{0.05cm}\includegraphics[scale=0.53]{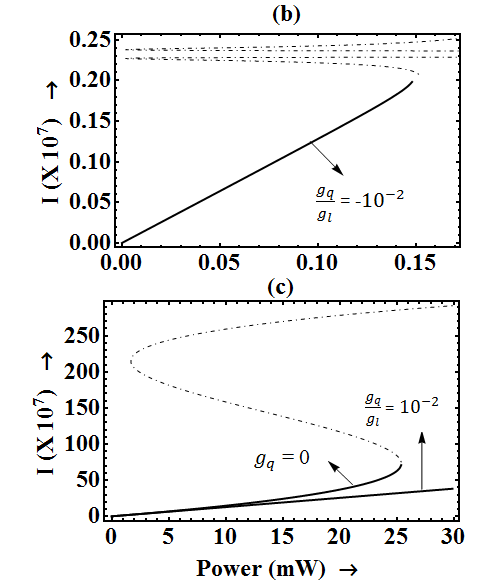}
  \caption{\label{fig4} (a) Shows the effective normalised spring constant $\frac{\tilde{\mathcal{K}}}{ \mathcal{K}} = \frac{\tilde{\omega}_m}{\omega_m}$ as  a function of power and QOC. (b) and (c) Shows intensity ($I$) inside the cavity varied as a function of input power ($\mathcal{P}$) for various QOCs ($g_{_q}/g_{_l}$) as shown. The solid line depicts the stable region and the dot-dashed line corresponds to unstable region. The system parameters are $\Delta=\omega_m$ and $g_{_{l}}/2\pi=215$ Hz at bath temperature $T=1$ mK.}
\end{figure}

The presence of QOC modifies both mechanical spring constant from $\mathcal{K}=m\omega_m^2$ to $\mathcal{\tilde{K}}=m\omega_m\tilde{\omega}_m$, as estimated from Eqn.(\ref{s1}b)\footnote{Note that Eqn.(\ref{s1}(b)) is expressed in dimensionless quantities. We convert $\mathbf{x,p}$ to $\mathbf{\hat{x},\hat{p}}$, as shown later in Sec.5 to deduce $\mathcal{K}$}. Therefore the normalized modified spring constant, which we define as $\mathcal{\tilde{K}}/\mathcal{K}$ is  proportional to $\tilde{\omega}_m/\omega_m$.  Along with magnitude, the sign of QOC also has a major effect on spring constant. The positive (negative) QOC stiffens (softens) the oscillator in comparison to the case of LOC alone. This can be understood from Fig. \ref{fig4}(a) which shows the variation of modified normalised spring constant as a function of power and QOC. The blue color region depicts the softening of spring for  negative QOC and red color corresponds to stiffening of spring for positive QOC. For a cavity-optomechanical system to be stable, it is necessary that radiation pressure force and the mechanical restoring force balance each other. Since the change in spring constant leads to a change in restoring force, the radiation pressure force, given by $F_{rad}= (\hbar \omega_c/L)\langle a^\dagger a\rangle\propto\frac{\mathcal{P}}{\tilde{\Delta}^2+\kappa^2}$, has to modify accordingly to retain stability. We show this by calculating intensity inside the cavity at various input powers by setting various QOC values using Eqn.(\ref{a5}). We verified these values with the Routh-Hurwitz criteria given in Eqn.(\ref{a15}) and plotted as a function of input power $\mathcal{P}$ as shown in the Fig. \ref{fig4}(b,c).  We chose negative QOC value of $g_{_{q}}/g_{_{l}}=-10^{-2}$ in Fig. \ref{fig4}b and  Fig. \ref{fig4}c is plotted for zero and positive QOC values ($g_{_{q}}/g_{_{l}}=10^{-2}$). The curves with black solid lines are the stable points and the curves with dot dashed lines are the unstable points making it transparent that with changing QOC values from negative to positive, it is possible to find stable regions ranging from microwatts to milliwatts.  
 
\section{Theoretical description of mechanical oscillator's fluctuations}
Using the definition of Fourier transform, $\mathcal{F}(\omega)=\frac{1}{2\pi}\int_{-\infty}^\infty\mathcal{F}(t)e^{-i\omega t}dt$ and $[\mathcal{F}^\dagger(\omega)]^\dagger=\mathcal{F}(-\omega)$  in the Eqn.(\ref{a12}), the set of coupled differential equations form a simple system of linear equations in frequency.  Therefore after solving the matrix equation Eqn.(\ref{a12}) in frequency domain, we get 
\small
\begin{equation}
\delta x (\omega)=\frac{1}{D(\omega)}\lbrace X_a(\omega) a_{in}(\omega)+ X_{a^\dagger}(\omega) a_{in}^\dagger (\omega) -X_{\xi}(\omega)\xi(\omega) \rbrace ,\label{b0}
\end{equation}\normalsize where \small
\begin{subequations}
\begin{eqnarray}
&&D(\omega)= \left((\kappa -i \omega)^2+\tilde{\Delta}^2\right)\left( \omega^2+i \gamma_m \omega- \omega_m \tilde{\omega}_m)\right) \nonumber\\&&+2\tilde{G}^2I \tilde{\Delta}\omega_m ,\\
&&X_a(\omega)=\sqrt{2 \kappa} \omega_m \tilde{G}a_s^* (\kappa - i\omega - 
   i \tilde{\Delta}),\\
&&X_{a^\dagger}(\omega)= (X_a(-\omega))^\dagger, \\
&&X_\xi(\omega)=\omega_m ((\kappa - i \omega)^2 + \tilde{\Delta}^2).
\end{eqnarray}\label{b2}
\end{subequations}\normalsize 
 The terms proportional to $\tilde{G}$ and  $\xi(\omega)$ in Eqn.(\ref{b0}) describes the effect of the radiation pressure and  thermal noise on mirror's motion respectively.\hspace*{+0.15cm}In the case of no coupling with the
cavity field, $\delta x(\omega)= \omega_m\xi(\omega)/\left(\omega_m^2-\omega^2-i\gamma_m\omega\right)$ the movable mirror will make Brownian motion, whose susceptibility has a
Lorentzian shape centred at frequency $\omega_m$ with width $\gamma_m$. But now, both thermal noise and radiation pressure decide the susceptibility.
The spectrum of fluctuations in position of the movable
mirror is defined by\small
\begin{equation}
S_{xx}(\omega)=\frac{1}{4\pi}\int e^{-i(\omega+\Omega) t} \langle \delta x(\omega)\delta x(\Omega) +\delta x(\Omega)\delta x(\omega)\rangle d\Omega.\label{a16}
\end{equation}
\normalsize
To evaluate the spectrum, the required correlations of noise operators in frequency domain are given by \small
\begin{subequations}
\begin{eqnarray}
&&\langle \delta a_{in}(\omega)\delta a_{in}^\dagger(\omega')\rangle =2\pi\delta(\omega+\omega'),\\
&&\langle \xi(\omega)\xi(\omega')\rangle = 2\pi \frac{\omega\gamma_m}{\omega_m}\left[\coth\left(\frac{\hbar \omega}{2k_BT}\right)+1\right]\delta(\omega+\omega').
\end{eqnarray}\label{b4}\end{subequations}\normalsize
Substituting Eqn.(\ref{b0}) and Eqn.(\ref{b4}) in Eqn.(\ref{a16}), we obtain the spectrum of fluctuations in position of the movable mirror, 
\begin{equation}
S_{xx}(\omega)=|\chi_{eff}(\omega)|^2[S_{th}(\omega)+S_{rp}(\omega,\tilde{\Delta})]
\end{equation} where $S_{th}(\omega)$, $S_{rp}(\omega)$ are the thermal and radiation pressure noise spectra respectively and the effective mechanical oscillator susceptibility are given by  \small\begin{subequations}
\begin{eqnarray}
&&S_{th}(\omega)=\frac{\omega\gamma_m}{\omega_m}\coth\left( \frac{\hbar\omega}{2k_BT}\right),\\
&&S_{rp}(\omega)=\frac{2\tilde{G}^2I\kappa(\kappa^2+\omega^2+\tilde{\Delta}^2)}{\left( \kappa^2+(\omega-\tilde{\Delta})^2\right)\left( \kappa^2+(\omega+\tilde{\Delta})^2\right)},\\
&&\chi_{eff}(\omega)=\omega_m \left[ \left( \omega_m \tilde{\omega}_m-\omega^2-i \gamma_m \omega \right)- \frac{2\tilde{G}^2I \tilde{\Delta}\omega_m}{(\kappa -i \omega)^2+\tilde{\Delta}^2} \right]^{-1} .\label{a23}
\end{eqnarray} \label{b6}\end{subequations}\normalsize 
From Eqn.(\ref{a23}), the effective frequency and damping rate of the mechanical oscillator are
\small \begin{subequations} \begin{eqnarray}
\Omega_{eff}(\omega)&=& \left[ \omega_m\tilde{\omega}_m-\frac{2\tilde{G}^2I \tilde{\Delta}\omega_m(\kappa^2-\omega^2+\tilde{\Delta}^2)}{\left( \kappa^2+(\omega-\tilde{\Delta})^2\right)\left( \kappa^2+(\omega+\tilde{\Delta})^2\right)}\right]^\frac{1}{2} \label{a24}\\
\Gamma_{eff}(\omega)&=&\gamma_m+\frac{4\kappa\tilde{G}^2I \tilde{\Delta}\omega_m}{\left( \kappa^2+(\omega-\tilde{\Delta})^2\right)\left( \kappa^2+(\omega+\tilde{\Delta})^2\right)} \label{a25}.
\end{eqnarray}\end{subequations}
\normalsize
The modification of mechanical frequency due to radiation pressure in Eqn.(\ref{a24}) is also known as the optical-spring effect.Then Fourier transforming Eqn.(\ref{a12}) for momentum fluctuations, we obtain $\delta p(\omega)=-i\omega/\omega_m\delta x (\omega)$, which can be used to calculate fluctuation spectrum of momentum as \small \begin{equation}
S_{pp}(\omega)=\frac{\omega^2}{\omega_m^2}S_{xx} (\omega). \label{a22}\end{equation} \normalsize

\section{\label{sec}Standard Quantum Limit (SQL) and squeezing of the mechanical motion}
Unlike a one-dimensional classical harmonic oscillator, the quantum harmonic oscillator has non-zero fluctuations in its ground-state due to the zero-point energy. These fluctuations, when calculated in the ground state come out to be $\sqrt{\hbar/2m \omega_m}$ and $\sqrt{\hbar m\omega_m/2}$ in its position ($\mathbf{\hat{x}}$) and momentum ($\mathbf{\hat{p}}$) respectively. These are known as the 'Standard Quantum Limit' (SQL). The presence of these non-zero fluctuations limit the ultra-high sensitivity of the optomechanical interferometers. But since the quadratures also satisfy the uncertainty principle $\Delta x \Delta p \geq \frac{1}{2}\bigg|\bigg\langle [\mathbf{\hat{x}},\mathbf{\hat{p}}]\bigg\rangle\bigg|$, we can only beat SQL by reducing fluctuations in one of the quadratures. Then the oscillator is said to be squeezed in its motion. 


Here we define the dimensionless position and momentum as $\mathbf{x}\equiv \mathbf{\hat{x}}\sqrt{m \omega_m/\hbar}$, $ \mathbf{p}\equiv \mathbf{\hat{p}} \sqrt{1/\hbar m \omega_m}$ ($\mathbf{\hat{x},}\mathbf{\hat{p}}$ are operators with dimensions and $\mathbf{x,p}$ are dimensionless operators).  
The variances of the mechanical oscillator's position and momentum can be evaluated from Eqn.(\ref{b6})-(\ref{a22}) as
\small
\begin{subequations}
\begin{eqnarray}
&&(\Delta x)^2=\frac{1}{2\pi}\int_{-\infty}^{\infty}S_{xx}(\omega)d\omega, \\
&&(\Delta p)^2=\frac{1}{2\pi}\int_{-\infty}^{\infty}S_{pp}(\omega)d\omega=\frac{1}{2\pi}\int_{-\infty}^{\infty}\frac{\omega^2}{\omega_m^2}S_{xx}(\omega)d\omega.
\end{eqnarray}\label{a21}
\end{subequations}\normalsize The variances for these dimensionless position and momentum calculated in the ground state has the values $(\Delta x)^2=(\Delta p)^2=1/2$ and this is defined as the SQL. When either of the variances is less than 1/2, SQL is said to be beaten and the corresponding quadrature being squeezed.

\section{Effect of QOC on squeezing the mechanical motion}

The presence of QOC along with LOC modifies the natural spring constant and thereby the resonance frequency of mechanical oscillator $\omega_m$ to its intensity dependent counterpart $\sqrt{\omega_m\tilde{\omega}_m}$. Also the single optomechanical coupling rate modifies $g_{_{l}}$ to $\tilde{G}$ driving the effective optomechanical interaction rate at $\tilde{G}I$. Thus, the system can be regarded as a cavity-field-driven parametric oscillator modifying the mechanical oscillator frequency. By choosing the strong coupling limit \textit{i.e.} $\tilde{G}I>\omega_m >\kappa$ not only enables us to study the system's steady-state behaviour, but also the modification of mechanical oscillator's frequency that occurs faster than its natural time-scale $1/\omega_m$. It has been shown that such a modification of harmonic oscillator's frequency in its coherent state affects the position and momentum quadratures \cite{graham}. Hence, unlike the conventional optomechanical systems (LOC alone), such a  system in its ground state can generate squeezed states. Further, these effects on quadratures are proportional to the modifications in the frequency.

In order to understand the effect of modification of mechanical frequency and other parameters dependence on squeezing, we simplify the integrals of Eqn.(\ref{a21}). We make the approximations i) of thermal noise contribution $S_{th}$ in Eqn. (\ref{b6}a) as $\frac{\omega\gamma_m}{\omega_m}\coth\left( \frac{\hbar\omega}{2k_BT}\right) \approx \frac{2\gamma_mk_BT}{\hbar\omega_m}=\gamma_m(2n_{th}+1)$  and ii) the radiation pressure contribution $S_{rp}$ in Eqn. (\ref{b6}b) as shown below by considering system to be in  quasiresonance regime. This enables us to consider  $\Gamma_{eff}(\omega)  \simeq \Gamma_{eff} (\sqrt{\omega_m \tilde{\omega}_m })$  and $\Omega_{eff}(\omega) \simeq \Omega_{eff} (\sqrt{(\omega_m\tilde{\omega}_m})$ giving us 
\small
\begin{eqnarray}
&&S_{xx}(\omega)=|\chi_{eff}(\omega)|^2\left(\gamma_m (2n_{th}+1)+ \frac{(\Gamma_{eff}-\gamma_m)(\kappa^2+\tilde{\Delta}^2+\omega^2)}{\omega_m\tilde{\Delta}}\right) \Biggg\vert_{\text{quasiresonance }}.\end{eqnarray}\normalsize
Finally we use cauchy-residues theorem to evaluate the integrals in Eqn. (\ref{a21}) giving us the variances. It is clear from these equations that the effective frequency $\Omega_{eff}$, which has a value around $\sqrt{\tilde{\omega}_m\omega_m}$ plays a major role in suppressing the variances.
\begin{subequations}
\begin{eqnarray}
&&(\Delta x)^2=\dfrac{\omega_m^2}{4\Omega_{eff}^2\Gamma_{eff}}\left(\gamma_m (2n_{th}+1)+\frac{\left(\Gamma_{eff}-\gamma_m\right)\left(\kappa^2+\tilde{\Delta}^2+\Omega_{eff}^2\right)}{\omega_m\tilde{\Delta}}\right),\\
&&(\Delta p)^2=\dfrac{1}{4\Gamma_{eff}}\left(\gamma_m (2n_{th}+1)+\frac{\left(\Gamma_{eff}-\gamma_m\right)\left(\kappa^2+\tilde{\Delta}^2+\Omega_{eff}^2-\Gamma_{eff}^2\right)}{\omega_m\tilde{\Delta}}\right).
\end{eqnarray}\label{squ} \end{subequations}\normalsize

\begin{figure*}[b]
\centering
\includegraphics[scale=0.5]{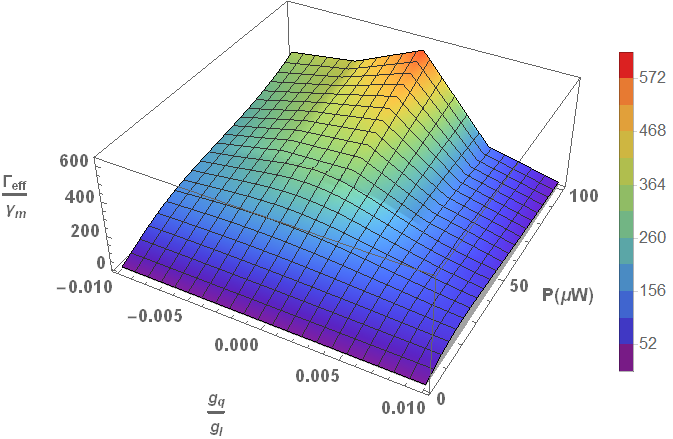}
\caption{\label{fig5}Shows effective damping rate $\Gamma_{eff}/\gamma_m$ evaluated at $\sqrt{\tilde{\omega}_m\omega_m}$, varied as a function of input power $\mathcal{P}$ and QOC ($g_{_{q}}/g_{_{l}}$). The parameters chosen are same as defined in Fig.\ref{fig4}.}
\end{figure*}

\begin{figure*}[h!]
\centering 
\hspace*{-0.8cm}\includegraphics[scale=0.3]{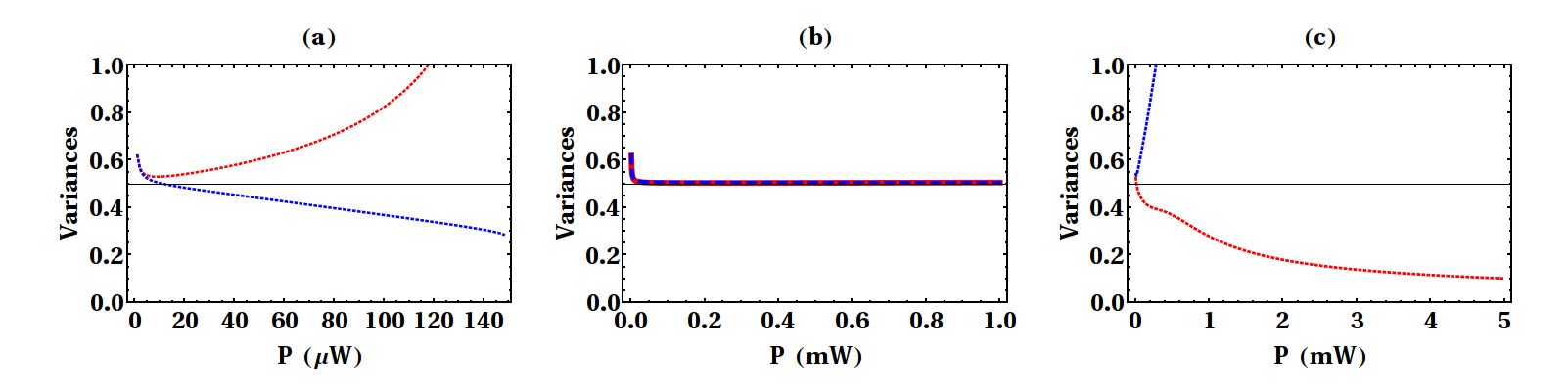}
\caption{\label{fig3} Shows the variances calculated numerically in position ($(\Delta x)^2$ as red color curve) and momentum ($(\Delta p)^2$ as blue color by  curve) using Eqn.(\ref{a21}), plotted as a function of  pump power ($\mathcal{P}$). The black solid line at 0.5 depicts SQL. The system parameters are same as that in Fig. \ref{fig4} with $g_{_{q}}/g_{_{l}}$ values varied as (a) $-10^{-2}$, (b) 0 and (c) $10^{-2}$.}
\end{figure*}

With the inclusion of QOC, not only the effective frequency of the system changes to around $\sqrt{\omega_m\tilde{\omega}_m}$, but also modifies its damping rate $\Gamma_{eff}$. The presence of damping rate (as a function of response frequency $\omega$) is attributed to the back-action effects of radiation pressure onto the mechanical oscillator. These backaction effects usually hinder the  suppression of quantum fluctuations below SQL in the case of conventional optomechanical systems (LOC alone, see Fig. \ref{fig3}(b)).  The advantage of adding a QOC interaction can be seen from Eqn. (\ref{a25}) that it decreases this back-action effect on the oscillator. This  is evident from Fig. \ref{fig5}, where $\Gamma_{eff}(\sqrt{\omega_m\tilde{\omega}_m})$ (considering the quasiresonant approximation) is plotted as a function of QOC and power. Here the effective damping rate at $g_{_{q}}/g_{_{l}}=-10^{-2}$ is half in comparison with the conventional system and is as less as 20 times for $g_{_{q}}/g_{_{l}}=10^{-2}$  at the same power of 100 $\mu$W. This striking feature of QOC on reduction in back-action effect and modification of mechanical frequency together aid in suppressing fluctuations below the SQL, provided an enough amount of radiation pressure is present. Therefore with an appropriate intensity $I$ (proportional to driving power $\mathcal{P}$), the quadratures can be squeezed. Using Eqn.(\ref{a21}) we calculated variances numerically that are shown in Fig. \ref{fig3} with red curves representing variances in position and blue curves representing variances in momentum plotted as a function of input power ($\mathcal{P}$), for various QOCs. The black solid line depicts SQL and the value of variance in a quadrature below this line represents squeezing of the corresponding quadrature. From Fig.\ref{fig3}(b) it is straightforward to see that, we find the quadratures at most reach only SQL with LOC alone and squeezing is observed in the case of Fig.\ref{fig3}(a),(c). Our analytical expression Eqn. (\ref{squ}) is in an excellent agreement with the numerical results plotted in Fig. \ref{fig3}.

It is clear from Eqn. (\ref{squ}) that the lower the $\Omega_{eff}$, lower the variance in momentum and vice versa. This is quite in agreement with \cite{graham}, in which it has been shown that the modified frequency being lesser than the natural frequency generates squeezing in momentum and vice versa. Therefore if QOC is chosen negative, then $\sqrt{\tilde{\omega}_m\omega_m}<\omega_m$ thus producing squeezed states in its momentum and  conversely for position. Hence choosing negative (positive) QOC in Fig. \ref{fig3}(a) (Fig. \ref{fig3}(c)), gives rise to a maximum squeezing in momentum  (position). Since there lies no more stable regions for negative QOC after $\sim$150 $\mu$W, the squeezing gets maximised and limited to nearly 40\% in momentum.  Whereas the availability of large stability range together with enormous reduction of back-action effects with positive value of QOC enables squeezing the position quadrature beyond the 3 dB limit ($>50\%$ below the SQL). This is shown in Fig. \ref{fig3}(c) where 3 dB limit is surpassed at 1.3 mW and further squeezing can be achieved by increasing the input power. 

It has been found theoretically that QOC alone can be used to a back-actionless trapping \cite{trapping} and cooling \cite{girvin} of the mechanical mirror and also anticipated in exploring other quantum features such as mechanical squeezing \cite{bhattacharya}, quantum jumps and  quantization of mechanical energy \cite{thompson}, photon transport \cite{photonshuttle} etc. very effectively. However instead of QOC alone, an optomechanical system with both LOC and residual QOC used together can explore  these quantum features very efficiently  and can reveal many more interesting features which would be a part of our future work.

\section{Conclusion}
We have studied and analysed a cavity optomechanical system in a resolved side-band regime, where dispersive linear and quadratic couplings together were considered. We showed that in presence of a residual QOC, as low as hundred times that of the already present LOC, back-action effects reduces enormously and a sudden change of mechanical oscillator's frequency occurs. This enables the system to beat the standard quantum limit. Such systems can be realised in tunable optomechanical systems like membrane in the middle set-up \cite{sankey} and atom-chip based systems \cite{purdy}.  Recent experimental progress has witnessed the presence of both LOC and QOC in various optomechanical systems \cite{doolin,flowers,barker}. In addition, the sign of quadratic coupling present in the system determines the quadrature in which squeezing occurs.  We further showed that QOC affects the stability range which could be utilised for squeezing the position quadrature of mechanical oscillator beyond the 3 dB limit. We look upon this as a pertinent result towards achieving the long-standing goal of ultra-sensitive measurements of weak forces.


\bibliographystyle{tMOP}
\bibliography{ref}

\end{document}